\documentclass[aps, pre, twocolumn, groupaddress, floatfix]{revtex4-1}
\usepackage{amsmath,amssymb,graphicx,color,soul}
\usepackage{epstopdf}

\newcommand{\TTS}{ {\bar t}_N }

\newcommand{\rT}{{\bf r}_{0}}

\definecolor{armygreen}{rgb}{0.29, 0.33, 0.13}
\definecolor{britishracinggreen}{rgb}{0.0, 0.26, 0.15}
\definecolor{cadet}{rgb}{0.33, 0.41, 0.47}
\definecolor{camouflagegreen}{rgb}{0.47, 0.53, 0.42}
\definecolor{cadmiumgreen}{rgb}{0.0, 0.42, 0.24}
\definecolor{darkgreen}{rgb}{0.0, 0.2, 0.13}
\definecolor{darkmidnightblue}{rgb}{0.0, 0.2, 0.4}
\definecolor{gray}{rgb}{0.7, 0.7, 0.7}

\begin{document}
	\title{Optimal Searcher Distribution for Parallel Random Target Searches}
	
	\author{Sunghan Ro}
	\address{Department of Physics, Technion-Israel Institute of Technology, Haifa 3200003, Israel}
	\author{Yong Woon Kim}
	\address{Department of Physics, Korea Advanced Institute of Science and Technology, Deajeon 34141, Korea}
	\address{Department of Physics, Massachusetts Institute of Technology, Cambridge, Massachusetts 02139, USA}
	
	\begin{abstract}	
		We consider a problem of finding a target located in a finite $d$-dimensional domain, using $N$ independent random walkers, when partial information on the target location is given as a probability distribution. When $N$ is large, the first-passage time sensitively depends on the initial searcher distribution, which invokes the question of what is the optimal searcher distribution that minimizes the first-passage time. Here, we analytically derive the equation for the optimal distribution and explore its limiting expressions. If the target volume can be ignored, the optimal distribution is proportional to the target distribution to the power of one-third. 
		If we consider a target of a finite volume and the probability of initial overlapping of searchers with the target cannot be ignored in the large $N$ limit, the optimal distribution has a weak dependence on the target distribution, given as a logarithm of the target distribution.
		Using Langevin dynamics simulations, we numerically demonstrate our predictions in one- and two-dimensions. 
	\end{abstract}
	
	
	\maketitle
	
	\section{Introduction}

	Target search by random walkers is a generic framework for studying numerous first-passage processes in nature, including chemical reaction, animal foraging, and genetic drift, to name a few~\cite{redner1984kinetics,redner2001guide,bramson_asymptotic_1988,pinsky_asymptotics_2003,moreau_lattice_2004,noh_random_2004,condamin_first-passage_2005,condamin_first-passage_2007,benichou_narrow-escape_2008,loverdo_enhanced_2008,oshanin_survival_2009,isaacson_reaction-diffusion_2009,benichou_geometry-controlled_2010,chevalier_first-passage_2011,isaacson_uniform_2013,bressloff_stochastic_2013,benichou_first-passage_2014,vaccario_first-passage_2015}. In many cases of search including animal foraging or finding a missing child, reducing the search time is highly desired. This motivated studies on the optimal walk statistics minimizing the search time, with lots of attention paid to L{\' e}vy flight~\cite{viswanathan_levy_2002,palyulin2014levy,levernier2020inverse}, intermittent search~\cite{o1990search,benichou_optimal_2005,benichou_intermittent_2011,volpe2017topography}, and more recently, non-Markovian search~\cite{meyer2021optimal}.
	
	Another way to reduce the search time is to deploy many searchers into the system~\cite{lindenberg_lattice_1980,krapivsky_kinetics_1996,yuste_order_1996,drager_sorting_1999,drager_mean_2000,yuste_diffusion_2000,rojo_lifetime_2010,mejia-monasterio_first_2011,meerson_mortality_2015,agranov_survival_2016,ro2017parallel,agranov_narrow_2018,lawley2020probabilistic,lawley2020universal,lawley_distribution_2020,lawley_probabilistic_2020,grebenkov2020single}. When $N$ searchers are looking for a target in parallel, the search time is determined by the first searcher to reach the target among them, and naturally, the search time is decreasing function of $N$. 
	If $N$ is small, the search time exhibits a universal behavior, inversely proportional to $N$, regardless of their initial distributions in the search domain~\cite{ro2017parallel}. As $N$ increases, however, the $N$ dependence of the search time shows drastic differences depending on the initial searcher distribution.
	If searchers of large $N$ are uniformly distributed, the search time becomes inversely proportional to $N^2$~\cite{ro2017parallel,grebenkov2020single}. 
	On the other hand, if the searchers depart from a point in the domain, the $N$ dependence of the search time shows much richer trends. In the intermediate value of $N$, the search time decreases exponentially with $N$ until it reaches the time required for a random walker to diffuse over the distance between its initial position and the target position. 
	As increasing $N$ further, the search time barely changes with $N$, entering into a regime of a weak logarithmic $N$ dependence~\cite{meerson_mortality_2015,ro2017parallel}. 

	Due to the sensitive dependency of the search time on the initial searcher distribution at large $N$, if one tries to minimize the search time, the information about the target position is crucial for deciding how to deploy the searchers. For example, let us consider the two limiting cases where the searchers are deployed either uniformly or at a point.
	If the target position can be specified with precision, then it is better to concentrate the searchers around the expected target location. On the other hand, if the target position is unknown, it is disadvantageous to place the searchers at a point, since the chosen location is unlikely to be the target position, and the search time would decrease only inversely proportional to $\ln N$. In contrast, the search time can be reduced by the factor of $N^{-2}$ if the searchers are distributed uniformly~\cite{ro2017parallel}. 
	
		\begin{figure} [b!]
		\center
		\includegraphics[width=1.0\linewidth]{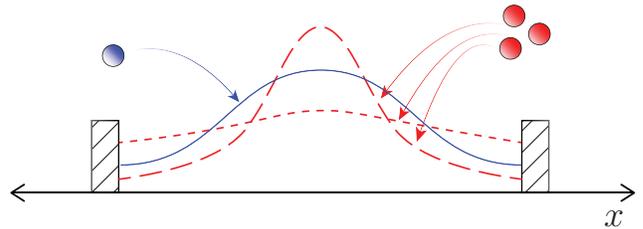}
		\caption{ 
		Schematic picture of a random target search by multiple searchers as the partial information of target position is given.
		When a target (blue ball) is located according to a probability distribution (blue line), various distributions of initial searcher positions (red lines) can be considered to minimize the search time.
		}
		\label{Fig:schematic}
		\end{figure}
	
	In this study, we extend this example to consider general searcher distributions, and ask the following question: What would be the optimal initial searcher distribution that minimizes the search time, if partial information on the target location is available in the form of a probability density function? (see Fig.~\ref{Fig:schematic} for illustration)
	To answer this question, we first derive the equation of the search time in the large $N$ limit when the target position and the initial searcher positions are given by arbitrary distributions. Then we put forward the way to determine the optimal searcher distribution when the target distribution is given.

	
	Our paper is organized as follows. In Sec.~\ref{Sec:system}, we define the system and introduce key quantities. 
	The main results of the paper are presented in Sec.~\ref{Sec:results}. 
	When the target volume can be neglected, the optimal distribution is shown to be proportional to the one-third power of the target distribution.
	When the target volume cannot be neglected, the optimal distribution has a weaker dependence, with its variation being proportional to the logarithm of the target distribution.
	Using the Langevin dynamics simulations, we also confirm these analytic predictions.
	In Sec.~\ref{Sec:conclusion}, we discuss the results and conclude.

	\section{System} \label{Sec:system}
	
	Consider a finite domain ${\cal D}$ with volume $V$ containing a small target located at ${\bf r}_0$, occupying a region of volume $V_T$ denoted as ${\cal T} ({\bf r}_0)$. The target volume is assumed to be very small compared to the domain volume, i.e., $V_\mathrm{T} \ll V$. Then $N$ searchers are introduced into the domain with initial positions specified by an array of vectors ${\cal R} = ({\bf r}_1, {\bf r}_2, \cdots, {\bf r}_N)$ where ${\bf r}_i$ is the initial position of the $i$-th searcher. 
	The searchers perform Brownian motion and randomly explore the domain in parallel. Once one of the searchers arrives at the target for the first time, the search process is ended and the time until the event is called the search time. 
	
	As defined above, the search time is a random variable. To analyze it, we first consider the probability density function (PDF) of searchers, which satisfies the diffusion equation given as
	\begin{equation} \label{Eq:Diff}
		\frac{\partial p\left({\bf r},t;  \rT; {\bf r}_{i} \right)}{\partial t} = D \nabla^2 p \left({\bf r},t; \rT; {\bf r}_{i}  \right),
	\end{equation}
	where the Laplace operator is applied to the searcher position ${\bf r}$ and $p({\bf r}, t; {\bf r}_0; {\bf r}_i)$ indicates the probability for the $i$-th searcher departed from ${\bf r}_i$ to be at ${\bf r}$ at time $t$, when the target is located at ${\bf r}_0$. The domain is surrounded by a hardwall boundary preventing the searchers from escaping, which is considered with a reflecting boundary condition $\hat{\bf n} \cdot \nabla p( {\bf r}, t; {\bf r}_0; {\bf r}_i)=0$ when ${\bf r} \in \partial{\cal D}$. Here, $\hat{\bf n}$ denotes the vector perpendicular to the domain boundary. Once the searchers arrive at the target, they are absorbed to the target, which imposes an absorbing boundary condition, $p( {\bf r}, t; {\bf r}_0; {\bf r}_i)=0$ when  ${\bf r} \in {\cal T}$ $({\bf r}_0)$. 
	
	The probability for each searcher to arrive at the target increases with time, leading to a decrease in the probability of remaining in the domain. For quantitative analysis, we define the survival probability of a searcher given as
	\begin{equation} \label{Eq:S}
		S(t;{\bf r}_0; {\bf r}_i) \equiv \int_{ {\cal D}^\ast } \mathrm{d}^d {\bf r} ~ p({\bf r}, t; {\bf r}_0; {\bf r}_i).
	\end{equation}
	where ${\cal D}^\ast = {\cal D} \setminus {\cal T}({\bf r}_0)$ is the domain outside the target. 
	Since the survival probability decreases as the probability for the searcher to find the target increases, the decreasing rate of the survival probability gives the first-passage probability of the searcher:
	\begin{equation}
		F(t;{\bf r}_0; {\bf r}_i) = - \frac{\partial}{\partial t} S(t; {\bf r}_0; {\bf r}_i)~.
	\end{equation}

	It is well-known that the first-passage probability for the collective search of $N$ searchers can be found by applying the order statistics on the survival probability of single searcher. With this approach, the distribution for the minimum of the first-passage times is given as 
	\begin{equation}
		\label{Eq:FN} F_N \left(t; {\bf r}_0; {\cal R} \right) = \sum_{i=1}^{N} F( t; {\bf r}_0; {\bf r}_i) \prod_{j \neq i} S(t; {\bf r}_0; {\bf r}_j ).
	\end{equation}
	Equation.~\eqref{Eq:FN} has a simple interpretation: The probability that the target is found for the first time at time $t$ is equal to the probability for one of the searchers to reach the target at $t$ while the other searchers have not arrived yet. 
	Finally, the search time is calculated by taking average with $F_N \left(t; {\bf r}_0; {\cal R} \right)$ as
	\begin{eqnarray}
		\nonumber t_N({\bf r}_0; {\cal R}) &\equiv& \int_0^\infty \mathrm{d} t ~ t F_N(t; {\bf r}_0; {\cal R}) \\
		&=& \int_0^\infty \mathrm{d} t ~ \prod_{i=1}^{N} S(t; {\bf r}_0; {\bf r}_i)~, \label{Eq:tN_all}
	\end{eqnarray}
	where the second equality is obtained with the integration by parts. 
	
	Next, following the problem setup we consider, we introduce the search times when the the initial positions of a target and of searchers are random variables. By denoting the PDF of the target as $P_\mathrm{T}({\bf r}_0)$ and the PDF of the searchers as $P_\mathrm{S}({\cal R})$, the average search time can be obtained as 
	\begin{eqnarray} \label{Eq:TTS}
		\TTS \equiv \int_{\cal D} \mathrm{d}^d {\bf r}_0 \int \mathrm{d} {\cal R}\, P_\mathrm{T}({\bf r}_0) P_\mathrm{S} ({\cal R}) t_N({\bf r}_0; {\cal R})~,
	\end{eqnarray}
	which is the key quantity of this study. 
	In this expression, integration over the target position is performed over the whole domain ${\cal D}$ and $\mathrm{d}{\cal R}$ indicates integral over the initial positions of searchers.
	In the following section, we derive the equation for the optimal searcher distribution based on the quantities introduced so far. 
	
	\section{Results} \label{Sec:results}
	
	To start with, let us briefly remind the readers of the setup. We consider a target located at a position which is a random variable following a given PDF $P_\mathrm{T}({\bf r}_0)$, and searchers {\it independently} deployed into the system following the searcher distribution we specify. Denoting the searcher distribution of an individual searcher as $u({\bf r})$, the full searcher distribution then satisfies $P_\mathrm{S} ({\cal R}) = \prod_{i=1}^N u({\bf r}_i)$. 
	Considering that the search time is significantly affected by the initial searcher position only when there are many searchers in the domain, we focus in the present study on the large $N$ limit, for which the search time follows the large $N$ asymptotic behavior everywhere in the system. 
	We allow the initial configuration where initial positions of searchers are overlapped with the target. 


	Next, consider a small target located at ${\bf r}_0$ and evaluate the search time recorded by $N$ searchers distributed according to $u({\bf r})$. Following Eq.~\eqref{Eq:TTS}, the search time can be written as
	\begin{equation} \label{Eq:SPA_to_TN}
		\TTS = \int_{\cal D} \mathrm{d}^d {\bf r}_0\, \bar{t}_{N, \mathrm{S}} ({\bf r}_0) P_\mathrm{T}( {\bf r}_0)
	\end{equation}
	where the searcher-position-averaged search time is defined as
	\begin{equation} \label{Eq:SPA}
		\bar{t}_{N, \mathrm{S}} ({\bf r}_0) = \int_0^\infty \mathrm{d} t\, \left[ \bar{S} (t; {\bf r}_0) \right]^N ~.
	\end{equation}
	Here, the average survival probability is also given as 
	\begin{equation} \label{Eq:S_average}
		\bar{S}(t;{\bf r}_0) \equiv \int_{\cal D} \mathrm{d}^d {\bf r} ~ S(t;{\bf r}_0, {\bf r}) u({\bf r})~.
	\end{equation}
	Since the initial overlapping with the target is allowed, it is possible that $S(t=0;{\bf r}_0, {\bf r}) = 0$ when ${\bf r} \in {\cal T}({\bf r}_0)$, which means that the target is found instantaneously.
	On the other hand, if ${\bf r} \in {\cal D}^\ast$, the survival probability satisfies $S(t=0;{\bf r}_0, {\bf r}) = 1$ at $t=0$ and gradually decreases with time.
    Thus, the average survival probability Eq.~\eqref{Eq:S_average} at $t=0$ satisfies
    \begin{equation} \label{Eq:S_av_initial}
        \bar{S}(0;{\bf r}_0) = 1 - \int_{{\cal T} ({\bf r}_0) } \mathrm{d}^d {\bf r} \, u({\bf r})~,
    \end{equation}
    and the decay of the survival probability is determined by trajectories of searchers in configurations where all searchers are initially placed outside the target.
    For large $N$ limit, $\bar{t}_{N, \mathrm{S}}$ is governed by the short-time behavior of the survival probability 
    due to the exponential dependence of the integrand of Eq.~\eqref{Eq:SPA} on $N$. 

	As Eqs.~\eqref{Eq:Diff} and \eqref{Eq:S} indicates, the survival probability of Eq.~\eqref{Eq:S_average} can be calculated by obtaining the searcher PDF that is evolving from the initial distribution $u({\bf r})$ according to the diffusion equation. This leads to 
	\begin{equation} \label{Eq:p_averaged}
		p({\bf r}, t;{\bf r}_0) = 
		\int_{{\cal D}^\ast} \mathrm{d}^d {\bf r}' \, G({\bf r}, t;{\bf r}_0;{\bf r}') u({\bf r}') ~,
	\end{equation}
	where $G({\bf r}, t;{\bf r}_0;{\bf r}')$ is the Green's function of the diffusion equation with an absorbing boundary condition at the target surface and a no current boundary condition at the domain boundaries. 
	
	To make further progress, we analyze the small $t$ behavior of Eq.~\eqref{Eq:p_averaged} and introduce approximations for both the Green's function and the searcher distribution. 
	For the sake of conciseness, we present the details of the approximations for a point target in a one-dimensional domain with the coordinate $x \in {\cal D}$ and target position $x_0$. The result can easily be extended to consider a target with a finite volume in higher dimensions as suggested later. 
	To proceed, it is useful to use the fact that in the large $N$ limit, the search time is mainly determined by the searchers that are initially placed very close to the target, which then arrive at the target within the time required to diffuse the distance from 
	the initial position of the searchers to the target surface. 
	A rough scale of this distance can be obtained by considering a thin layer around the target and then by calculating the thickness $d_s$ required for this layer to contain a searcher for the given searcher density.
	This leads to the following relation for $d_s$: 
	\begin{equation} \label{Eq:d_s}
		\int_{|x - x_0|<d_s} \mathrm{d} x ~ Nu(x) \sim 1.
	\end{equation}
	The relation indicates that $d_s$ would roughly be inversely proportional to $N$. Therefore, in the large $N$ limit, it is sufficient to consider dynamics of the searchers very close to the target, and the reflective boundary condition at the end of the search domain does not play a significant role. Taking this into an account, we simplify the Green's function by ignoring the domain boundary, while only considering the absorbing boundary at the target surface. Then the Green's function can be found using the image method as
	$G(x,t;x_0;x') = G_0(x-x',t) - G_0(x-2 x_0 + x',t)$,
	where $G_0(x,t) \equiv (4 \pi D t)^{-1/2} \exp \left[ - x^2/(4 Dt) \right]$ is the Gaussian kernel in the free space. 
	
	As a next step of the approximation, we expand the initial searcher distribution around the target location as $u(x') = u(x_0) + (\partial u/ \partial x)|_{x=x_0} (x'-x_0) + {\cal O} (|x'-x_0|^2)$. Then, we estimate the scale of contributions of each term in the expansion to the search time. By inserting the expansion into Eq.~\eqref{Eq:d_s}, we can estimate the number of searchers around the target captured by each term in the expansion. As a result, the scale of the number of searchers captured by the $n$-th order term is be estimated as ${\cal O}(d_s^{n+1})$. Therefore, in the large $N$ limit where $d_s$ becomes very small, it is sufficient to consider only the lowest order term of the searcher distribution, $u(x_0)$, to estimate the search time. 
	
	By combining the approximations on the Green's function and the searcher distribution, then inserting these into Eq.~\eqref{Eq:p_averaged}, we can analytically calculate the PDF of each searcher near by the target as
	\begin{equation}
		p(x, t' ;x_0)  \underset{x\sim x_0}{\simeq} \mathrm{sgn}(x-x_0) u(x_0) \mathrm{erf} \left( \frac{x - x_0}{\sqrt{4 D t}} \right)~,
	\end{equation}
	where $\mathrm{sgn}(x) = -1$ if $ x<0$, and $\mathrm{sgn}(x) =1$ if $x>0$. 
	By integrating the PDF over the spatial domain, we obtain the average survival probability given in Eq.~\eqref{Eq:S_average} as
	\begin{equation} \label{Eq:1d_s_approx}
		\bar{S}(t;x_0) \simeq 1 - 4 u(x_0) \sqrt{ \frac{D t}{\pi } }~.
	\end{equation}
	Inserting Eq.~\eqref{Eq:1d_s_approx} into Eq.~\eqref{Eq:SPA}, we obtain the asymptotic expression for the searcher-position-averaged search time 
	\begin{eqnarray}
		\nonumber \bar{t}_{N, \mathrm{S}} (x_0) &\simeq& \int_0^\infty \mathrm{d} t\, \left[ 1 - 4 u(x_0)  \sqrt{ {D t}/{\pi } } \right]^N \\
		\nonumber &\simeq& \int_0^\infty \mathrm{d} t\, \exp \left[ -4 N u(x_0) \sqrt{Dt / \pi} \right] \\
		\nonumber &=& \frac{\pi}{8 D N^2 u^2(x_0)}~.
	\end{eqnarray}
	Lastly, if the target has a finite volume so that the initial overlapping with searchers occurs, 
	the initial value of $\bar{S}(t;x_0)$ given by Eq.~\eqref{Eq:1d_s_approx} becomes less than 1 and is reduced by a factor of $\bar{S}(0; x_0)$ in Eq.~\eqref{Eq:S_av_initial}, which is approximated in the small target limit as $\bar{S}(0;x_0) \simeq 1 - V_\mathrm{T} u(x_0) $.
	Other than the normalization of the initial value, the decay of $\bar{S}(t;x_0)$ remains the same as in the zero target-volume limit
	since the first-passage dynamics for $t >0$ is governed by the trajectories of initially non-overlapping searchers.
	As a result, the search time satisfies
	\begin{equation}
	    \bar{t}_{N, \mathrm{S}}(x_0) \simeq [1 - V_\mathrm{T} u(x_0)]^N \frac{ \pi }{8 D N^2 u^2(x_0) }~.
	\end{equation}
	
	For higher dimensions, the SPA search time can be generalized as~\cite{grebenkov2020single}
	\begin{equation} \label{Eq:SPA_d_dim}
		\bar{t}_{N, \mathrm{S}} ({\bf r}_0) \simeq [1 - V_\mathrm{T} u({\bf r}_0)]^N  \frac{\pi }{2 D S_\mathrm{T}^2 N^2 u^2 ({\bf r}_0) }~,
	\end{equation}
	with the surface area $S_\mathrm{T}$ of the target. In one-dimension, we set $S_\mathrm{T} = 2$ since the target has two sides that can act as absorbing boundaries. 
	
	Next, we take the target distribution into an account and address the equation for the fully-averaged search time. To do so, we insert Eq.~\eqref{Eq:SPA_d_dim} into Eq.~\eqref{Eq:SPA_to_TN} to obtain the expression for the fully-averaged search time 
	\begin{equation} \label{Eq:TTSlargeN}
		\TTS = \frac{\pi}{2DS_\mathrm{T}^{2} N^2} \int_{{\cal D}} \mathrm{d}^d {\bf r}_0\, [1 - V_\mathrm{T} u({\bf r}_0)]^N \frac{P_\mathrm{T}({\bf r}_0)}{u^2 ({\bf r}_0)}~,
	\end{equation}
	which has the form of a functional of the searcher distribution $u({\bf r})$. 
	Analyzing Eq.~\eqref{Eq:TTSlargeN}, we derive the equations for the optimal searcher distribution. The analysis leads to two notable cases: i) When $V_\mathrm{T} \to 0$ limit can be taken while $S_\mathrm{T}$ remains finite. 
	ii) When finite $V_\mathrm{T}$ should be considered. We describe the result for each case in the following.
	
	\subsection{The case with $V_\mathrm{T} \to 0$ and a finite $S_\mathrm{T}$}
	
	The first case is the limit in which the target volume can be ignored while its surface area remains finite. Such limit can be attained when the system is one-dimensional for which $S_\mathrm{T} = 2$ regardless of the target size, or when the target has highly elongated shapes, such as a needle in a two-dimensional domain or a disk in a three-dimensional domain. For these cases, the integrand of Eq.~\eqref{Eq:TTSlargeN} can be simplified by setting $V_\mathrm{T} = 0$. Then, we calculate the variation of the 
	search time when the searcher distribution is varied as $u({\bf r}) \rightarrow u({\bf r}) + \epsilon h({\bf r})$ with a small parameter $\epsilon \ll 1$. For the resulting variation $\delta \TTS \equiv \TTS[u({\bf r}) + \epsilon h({\bf r})] - \TTS[u({\bf r})]$, we obtain
	\begin{equation}
		\nonumber \delta \TTS = - \frac{\pi}{D S_\mathrm{T}^2 N^2}
		\int_{\cal D} \mathrm{d}^d {\bf r}_0 ~ P_T({\bf r}_0) \frac{\epsilon h({\bf r}_0)}{u^3({\bf r}_0)} + {\cal O} (\epsilon^2 )~.
	\end{equation}
	Here, the function for variation $h({\bf r})$ must satisfy $\int_{{\cal D}} \mathrm{d}^d {\bf r} ~ h({\bf r}) = 0$ due to the normalization condition imposed on $u({\bf r})$. When the searcher distribution has its optimal form $u^\ast({\bf r})$, the term linear to $\epsilon$ in the variation $\delta \TTS$ should vanish. This condition is satisfied if $P_T({\bf r}_0) / [u^\ast({\bf r}_0)]^3$ is a constant, which leads to the equation for the optimal distribution given as
	\begin{equation} \label{Eq:u_small}
		u^\ast ({\bf r}) = {\cal N} P^{\frac{1}{3}}_\mathrm{T} ({\bf r})~.
	\end{equation}
	Here, ${\cal N}$ is the normalization constant calculated from $\int_{{\cal D}} \mathrm{d}^d {\bf r} ~ u^\ast({\bf r}) = 1$. 
	Remarkably, the optimal distribution for this case is independent of the number of searchers.
	This result suggests that it is beneficial to spread the searchers than concentrate them to minimize the search time. For example, if the target distribution has the shape of a Gaussian distribution with width $\sigma$, the search time is minimized when the searcher distribution is a Gaussian distribution with the width $\sqrt{3}\sigma$. 
	
	\begin{figure} [t]
		\center
		\includegraphics[width=1.0\linewidth]{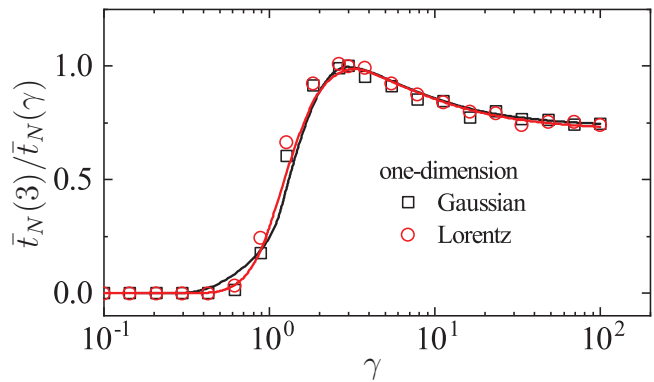}
		\caption{The inverse of the search times for various target and searcher distributions in one-dimension. 
			The symbols are obtained by averaging the search times recorded in the Langevin dynamics simulations over many realizations of the target and the searcher initial positions. The lines are analytic results obtained using Eq.~\eqref{Eq:TTSlargeN}.
			The parameters used are: $N = 10^4$, $\sigma = 0.35 L$ (Gaussian) and $\sigma = 0.2 L$ (Lorentz). The search times are obtained by taking averages over $10^3$ independent samples. The system size is $L=100a$ where $a$ is a unit length, and typical step size of the random walk is $0.01 a$. 
		}
		\label{Fig:largeN}
	\end{figure}
	
	To test the validity of this result, we perform Langevin dynamics simulations for the random search process and compare the search times as systematically varying searcher distribution. 
	In doing so, we consider a one-dimensional domain ${\cal D} = [-L,L]$ and a target with zero volume placed at a position $x_0$ which is sampled from a random initial distribution $P_\mathrm{T}({x}_0)$. 
	For distributions, we include a Gaussian distribution $P_\mathrm{T} (x_0) \propto \exp[-x_0^2/(2 \sigma^2)]$, and a Lorentz distribution $P_\mathrm{T}(x_0) \propto \sigma / (x_0^2 + \sigma^2)$, both of which are truncated at the domain boundaries.
	Then we introduce $N$ searchers randomly placed according to the searcher distribution given as $u_\gamma(x) = {\cal N}(\gamma) P_\mathrm{T}^{1/\gamma}(x)$ where ${\cal N}(\gamma)$ is the normalization constant for the given $\gamma$. For each realization, the search time is measured with the time recorded by the first searcher to reach the target, which is then averaged over many realizations of different positions of the target and searchers. We denote the resulting averaged searcher time as $\TTS(\gamma)$. 
	According to our result shown in Eq.~\eqref{Eq:u_small}, we expect the search time to have a minimum value at $\gamma = 3$, at which the searcher distribution becomes its optimal form. 
	In Fig.~\ref{Fig:largeN}, we plot normalized inverse search time $\TTS(3) / \TTS(\gamma)$ with respect to $\gamma$, for which we expect to observe a maximum at $\gamma = 3$. As shown in Fig.~\ref{Fig:largeN} this expectation is clearly confirmed from our simulation.

	\subsection{The case with a finite $V_\mathrm{T}$}
	
	Next, we consider the optimal distribution when $V_\mathrm{T}$ cannot be ignored. The procedure for obtaining the equation for the optimal distribution is the same as in the previous case, but now the functional variation has to be taken for the full expression of Eq.~\eqref{Eq:TTSlargeN}. The optimal distribution obtained may have a complicated form.
	As presented in Appendix.~\ref{App:derivation}, however, a rather simple expression can be obtained in the large $N$ limit, which reads as
	\begin{equation} \label{Eq:optimal_finiteVT}
	    u^\ast ({\bf r}) = c + \frac{1}{N V_\mathrm{T}} \ln [ P_\mathrm{T} ({\bf r})]~,
	\end{equation}
	where $c$ is an irrelevant constant determined by the normalization condition. 
	We note that 
	the optimal distribution is proportional to a logarithm of the target distribution.
	When the probability of the initial overlapping of searchers with a target is finite, it is better to distribute searchers so that they have a much wider distribution than $P_\mathrm{T} ({\bf r})$, as explicitly shown in the weak logarithmic dependence.
	Again, we test this prediction numerically by measuring the search time of $N$ searchers with the initial distribution given as $u_\gamma({\bf r}) = {\cal N}({\gamma}) P_\mathrm{T}^{1/\gamma}({\bf r})$.
	Although the optimal distribution for this case cannot be written as a power-law of the target distribution, $u_\gamma({\bf r})$ is still a fair approximation of it in the large $N$ limit as we argue below.
	
		\begin{figure} [t!]
		\center
		\includegraphics[width=1.0\linewidth]{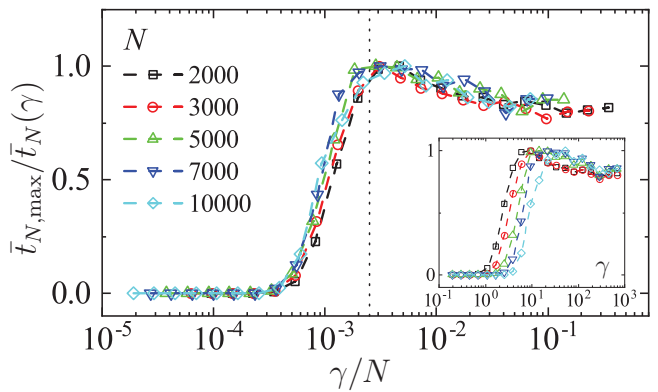}
		\caption{The inverse of the search times vs. $\gamma/N$ for a finite-sized target in a two-dimensional circular domain.
			The symbols indicate average search times measured from the Langevin dynamics simulations for various values of $N$. The dotted line indicates the value of $\gamma / N$ specified by Eq.~\eqref{Eq:gamma_N}. (inset) The inverse of the search times with respect to $\gamma$. 
			The parameters used are as follows: the target radius $a = 1$ (taken as a unit length), the domain radius $L = 20$, and $\sigma = 0.5 L$ for the target distribution. To obtain the average search time by $N$ searchers, we gathered search times of independent realizations of initial configurations, until $10^7 / N$ of non-zero search times are recorded. The typical step size of searchers per each time step is $0.01 a$. 
		}
		\label{Fig:collapse}
	\end{figure}
	
	To show this, we first rewrite the target distribution as $P_\mathrm{T} ({\bf r}) = A \exp [-I({\bf r})]$. Then the optimal distribution reads
	\begin{equation}
	    u^\ast ({\bf r}) = c' - \frac{1}{N V_\mathrm{T}} I({\bf r})~,
	\end{equation}
	where $c'$ is a constant. Accordingly, $u_\gamma({\bf r})$ can be written as
	\begin{equation} \label{Eq:u_gamma_I}
	    u_\gamma ({\bf r}) = {\cal N}(\gamma) \exp\left[ - \gamma^{-1} I({\bf r}) \right]~.
	\end{equation}
	When $\gamma$ is large, we expand Eq.~\eqref{Eq:u_gamma_I} in $\gamma^{-1}$ about $\gamma^{-1} = 0$ which leads to ${\cal N}(\gamma) = {\cal N}_0 + \gamma^{-1} {\cal N}_1 + \cdots$ and $\exp[-\gamma^{-1} I({\bf r})] \simeq 1 - \gamma^{-1} I({\bf r}) + \cdots$. By keeping terms of order up to $\gamma^{-1}$, Eq.~\eqref{Eq:u_gamma_I} can be written as
	\begin{equation}
	    u_\gamma ({\bf r}) \simeq ({\cal N}_0 + \gamma^{-1} {\cal N}_1) - {\cal N}_0 \gamma^{-1} I({\bf r}) ~.
	\end{equation}
	Therefore, $u_\gamma ({\bf r})$ with $\gamma = NV_\mathrm{T} {\cal N}_0$ converges to the optimal distribution in the large $N$ limit. 
	
	The analysis above shows that if the search time is plotted with respect to $\gamma$, it reaches its minimum when $\gamma = NV_\mathrm{T} {\cal N}_0$. Therefore, if the inverses of the search times obtained with various values of $N$ are plotted with respect to $\gamma / N$, we expect them to reach the maximum value all together when 
	\begin{equation} \label{Eq:gamma_N}
	    \frac{\gamma}{N} = \frac{V_\mathrm{T}}{ V}~,
	\end{equation}
	in the small $V_\mathrm{T}$ and the large $N$ limit. Here, we used the relation ${\cal N}_0 = V^{-1}$. We numerically test this expectation with random searches for a small but finite-sized target in two-dimensions. 
	To do so, we consider a circular domain ${\cal D}$ with a radius $L$ and a target of a radius $a$ randomly distributed according to the Gaussian distribution centered at the domain center, given as $P_\mathrm{T}({\bf r}_0) = {\cal N}_G \exp \left[ - |{\bf r}_0^2| / (2 \sigma^2) \right] $. Then we introduce the searchers according to the initial positions sampled with $u_\gamma ({\bf r})$, and measure the search time recorded by $N$ searchers. 
	Considering different $N$, the inverses of the search times obtained are plotted with respect to $\gamma$ in the inset of Fig.~\ref{Fig:collapse}. The inverse search times reach respective maxima at different values of $\gamma$. Then, when plotted with respect to $\gamma/N$ as shown in the main panel, these maximum points collapse with each other at $\gamma / N = V_\mathrm{T}/V$ as predicted by Eq.~\eqref{Eq:gamma_N}. Note that the location of $\gamma/N$ given as ~\eqref{Eq:gamma_N} is indicated with a dotted line in the figure, and the search times indeed reach their minimum regardless of the specific value of $N$ as predicted.

	\section{Discussion and conclusion} \label{Sec:conclusion}
	
	In this work, we have considered collective target search by many Brownian searchers when the target position is partially known in the form of a PDF $P_\mathrm{T}({\bf r})$, and have found the optimal distribution of the searchers which minimizes the search time. 
	To do so, we have introduced a systematic approximation for the survival probability of searchers and have expressed the search time as a functional of the searcher distribution. Examining the condition for the variation of the search time to vanish, we have derived equations for the optimal distribution in terms of the target distribution and the system parameters. In the large $N$ and the small $V_\mathrm{T}$ limit, the optimal distribution reaches one of two limiting forms depending on whether $V_\mathrm{T}$ can be ignored or not. 
	In most of previous studies, it was constrained that searchers are not initially overlapped with the target, corresponding to the $V_{\mathrm{T}} \rightarrow 0$ limit, which is released in the present study.
	For a target of a finite size, the initial overlapping configuration becomes non-negligible in the large $N$ limit, leading to distinct functional dependence of the optimal distribution on the target distribution.
	
	In the case when $V_\mathrm{T}$ can be ignored, e.g., when a point target is located in a one-dimensional search domain, the optimal distribution satisfies $u^\ast({\bf r}) \propto P_\mathrm{T}^{1/3}({\bf r})$ regardless of the number of searchers. 
	For the case where $V_\mathrm{T}$ cannot be ignored, the optimal distribution satisfies $u^\ast({\bf r}) \propto \ln [P_\mathrm{T} ({\bf r})] / (N V_T)$. 
	We have confirmed these predictions with the Langevin dynamic simulations. In particular, for the case when $V_\mathrm{T}$ cannot be ignored, we have found that the optimal distribution still approximately satisfies $u^\ast({\bf r}) \propto P_\mathrm{T}^{1/\gamma} ({\bf r})$ with $\gamma = N V_\mathrm{T} /V$, i.e., $\gamma$ is given as the number of searchers initially placed in the target volume when the searchers are uniformly distributed in the domain. 
	
	In this work, we have considered $N$ non-interacting searchers. Even in the absence of mutual interactions, the first-passage dynamics of multiple searchers can be complicated because the order statistics of $N$ passage times are required.
	It is recently shown that the presence of the interaction among searchers leads to non-trivial behaviors in the search time~\cite{choi2021first}. Naturally, the optimal searcher distribution may show sensitive dependence on the details of the searcher interaction as well. Thus, it would be a definitely interesting avenue to extend the current study to consider the situation where $N$ interacting searchers are involved.

	\acknowledgements
	This research was supported by a National Research Foundation of Korea (NRF) grant funded by the Korean government (Grant No. NRF-2020R1A2C1014826).
	
	\appendix
	
	\section{Derivation of Eq.~\eqref{Eq:optimal_finiteVT}} \label{App:derivation}
	To derive the equation, we consider variation of the search time $\delta \bar{t}_N = \bar{t}_N (\epsilon) - \bar{t}_N (0)$ with $\bar{t}_N$ given by Eq.~\eqref{Eq:TTSlargeN}, upon taking $u({\bf r}) \to u({\bf r}) + \epsilon h({\bf r})$. Here, $h({\bf r})$ is an arbitrary function satisfying $\int_{\cal D} \mathrm{d}^d \, h({\bf r}) = 0$. If $u({\bf r})$ is the optimal distribution, $\delta \bar{t}_N = 0$ should be satisfied, which leads to 
	\begin{equation}
	    P_\mathrm{T} ({\bf r}) \frac{[1 - V_\mathrm{T} u({\bf r})]^N}{ S_\mathrm{T}^2 N^2 u^3 ({\bf r})} \left( 1 + \frac{N V_\mathrm{T} u({\bf r})}{2 [ 1 - V_\mathrm{T} u({\bf r})]} \right) = c_0~, \nonumber
	\end{equation}
	where $c_0$ is a constant. To proceed, we use approximations 
	\begin{eqnarray}
	    \nonumber [1 - V_\mathrm{T} u({\bf r})]^N &\simeq& e^{-N V_\mathrm{T} u({\bf r})} \\
	    \nonumber \left( 1 + \frac{N V_\mathrm{T} u({\bf r})}{2 [ 1 - V_\mathrm{T} u({\bf r})]} \right) &\simeq& \frac{1}{2} N V_\mathrm{T} u({\bf r})~,
	\end{eqnarray}
	which are valid in the small $V_\mathrm{T}$ and the large $N$ limit, i.e.,  $V_T/V \ll 1$ and $N V_T / V \gg 1$.
	By rearranging the terms, we arrive at 
	\begin{equation}
	    \sqrt{P_\mathrm{T} ({\bf r}) \frac{N V_\mathrm{T}^3}{2 c_0 S_\mathrm{T}^2} } = N V_\mathrm{T} u({\bf r}) e^{N V_\mathrm{T} u({\bf r})/2}  \nonumber
	\end{equation}
	The optimal distribution can be obtained by solving this equation for $u({\bf r})$, which leads to 
	\begin{equation}
	    u^\ast ({\bf r}) = \frac{2}{N V_\mathrm{T}} W_{0} \left[ \frac{1}{2}  \sqrt{c_1 V_\mathrm{T} P_\mathrm{T} ({\bf r})} \right] ~,
	\end{equation}
	where $W_{0} (x) $ is the Lambert W function and $c_1 = NV_\mathrm{T}^2 /(2 c_0 S_\mathrm{T}^2)$ is a dimensionless number. For large values of $x$, this function can be approximated as $W_{0} (x) = \ln x + {\cal O}(\ln \ln x)$. Therefore, we obtain the asymptotic expression for the optimal searcher distribution in the large $N$ limit as
	\begin{equation}
	    u^\ast ({\bf r}) \simeq c + \frac{1}{N V_\mathrm{T}} \ln [V_\mathrm{T} P_\mathrm{T} ({\bf r})]~, \nonumber
	\end{equation}
	where $c$ is a constant determined by the normalization condition, and Eq.~\eqref{Eq:optimal_finiteVT} is derived. 
	
	
	%

\end{document}